\documentclass[aps,prl,twocolumn,showpacs,superscriptaddress,groupedaddress]{revtex4}  % for review and submission
\usepackage{graphicx}  % needed for figures
\usepackage{dcolumn}   % needed for some tables
\usepackage{bm}        % for math
\usepackage{amssymb}   % for math
\usepackage{color}
\usepackage[draft]{hyperref}
\usepackage{comment}
\usepackage[normalem]{ulem}
\usepackage{color,soul}

\begin{document}

% The following information is for internal review, please remove them for submission
%\widetext
%\leftline{Version 01 as of \today}
%\leftline{Primary author: Ben Chapman}
%\leftline{To be submitted to PRL}
%\leftline{Comment to {\tt B.Chapman@warwick.ac.uk} }
%\centerline{\em INTERNAL DOCUMENT -- NOT FOR PUBLIC DISTRIBUTION}

\title{Sub-microsecond temporal evolution of edge density during edge localized modes in KSTAR tokamak plasmas inferred from ion cyclotron emission}
\author{B. ~Chapman} \affiliation{Centre for Fusion, Space and Astrophysics, University of Warwick, Coventry, CV4 7AL, UK}
\author{R.O.~Dendy} \affiliation{CCFE, Culham Science Centre, Abingdon, OX14 3DB, UK}\affiliation{Centre for Fusion, Space and Astrophysics, University of Warwick, Coventry, CV4 7AL, UK}
\author{K.G.~McClements} \affiliation{CCFE, Culham Science Centre, Abingdon, OX14 3DB, UK}
\author{S.C.~Chapman} \affiliation{Centre for Fusion, Space and Astrophysics, University of Warwick, Coventry, CV4 7AL, UK}
\author{G.S.~Yun} \affiliation{Pohang University of Science and Technology, Pohang, Gyeongbuk 37673, Republic of Korea}
\author{S.G.~Thatipamula} \affiliation{Pohang University of Science and Technology, Pohang, Gyeongbuk 37673, Republic of Korea}
\author{M.H.~Kim} \affiliation{Pohang University of Science and Technology, Pohang, Gyeongbuk 37673, Republic of Korea}
\vskip 0.25cm

\date{\today}

\begin{abstract}
Ion cyclotron emission (ICE) is detected during edge localised modes (ELMs) in the KSTAR tokamak at harmonics of the proton cyclotron frequency in the outer plasma edge. The emission typically chirps downward (occasionally upward) during ELM crashes, and is driven by confined 3MeV fusion-born protons that have large drift excursions from the plasma core. We exploit fully kinetic simulations at multiple plasma densities to match the time-evolving features of the chirping ICE. This yields a unique, very high time resolution $\left(< 1 \mu s \right)$ diagnostic of the collapsing edge pedestal density.
\end{abstract}


\maketitle

\begin{figure}[ht]
\includegraphics[clip=true, trim = 0.0cm 0.0cm 0.0cm 0.0cm, width = \columnwidth]{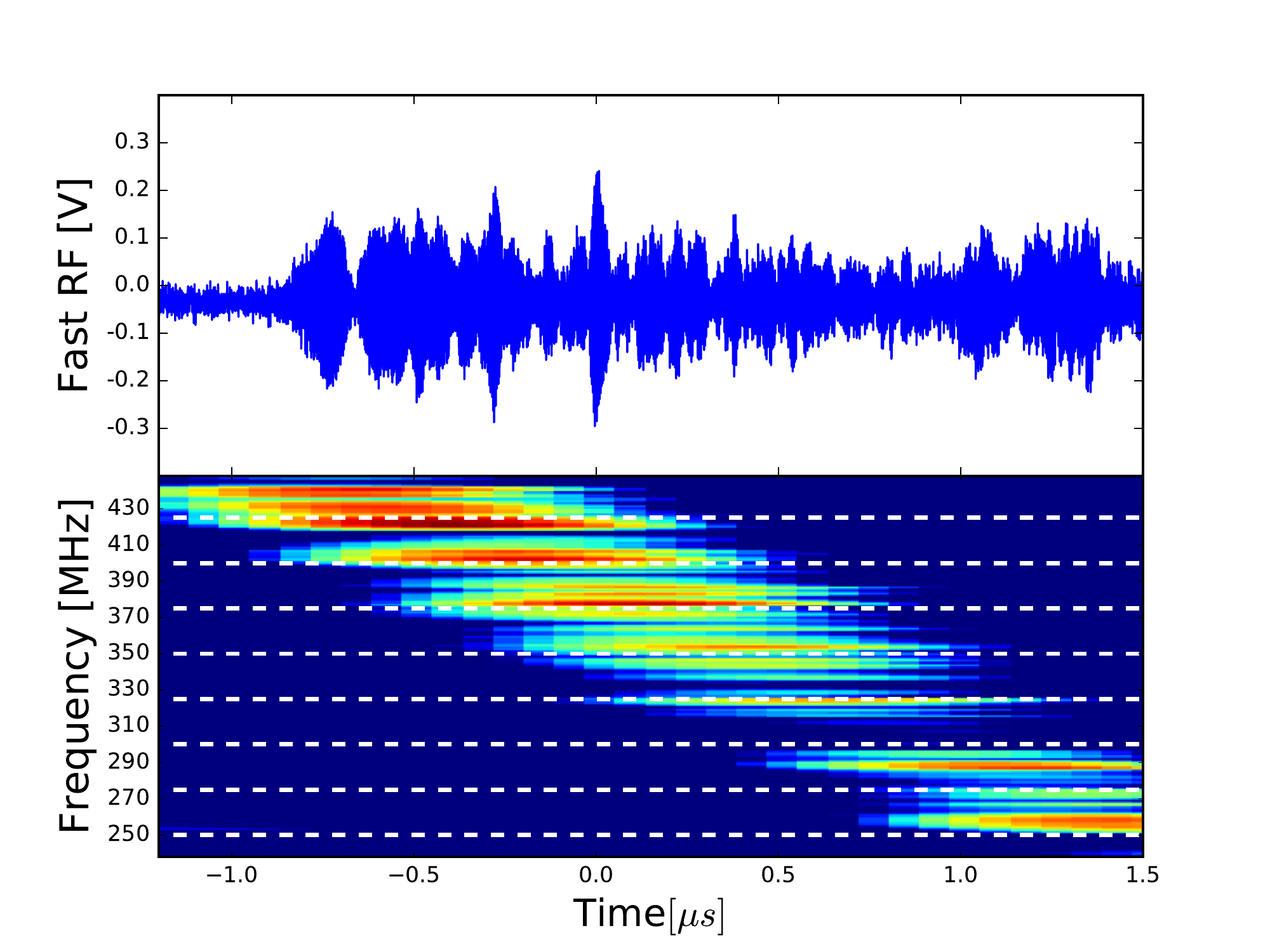}
\vspace{-0.8cm}
\caption{\label{fig:burstspec} Temporal evolution of ICE amplitude (upper plot) and spectrum (lower plot) during an ELM crash in KSTAR plasma 11513. Time is measured relative to the moment chirping bursts are observed during the ELM crash ($\sim 100 \mu s$ after the start of the crash). The horizontal dashed lines in the spectrogram indicate energetic proton cyclotron harmonics.}
\end{figure}

\begin{figure}[ht]
\includegraphics[clip=true, trim = 0.0cm 2.5cm 3.0cm 2.5cm, width = \columnwidth]{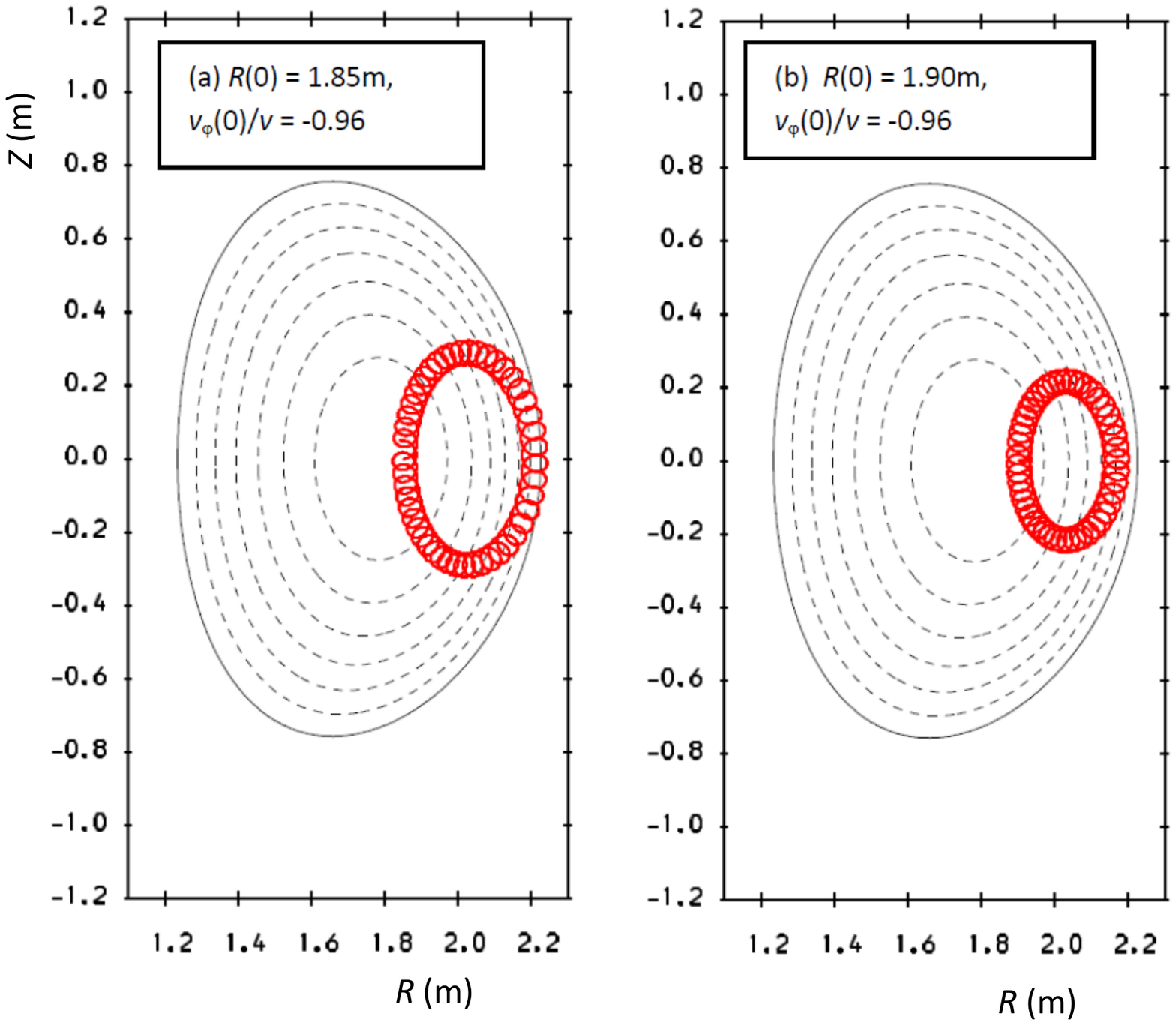}
\caption{\label{fig:orbits}  Poloidal projection of 3.0$\,$MeV fusion proton orbits in the model KSTAR equilibrium with initial velocity vectors slightly offset from the co-current toroidal direction.}
\end{figure}

\begin{figure}[!htb]
  \includegraphics[clip=true, trim = 0.5cm 0.5cm 1.0cm 1.0cm, width = \columnwidth]{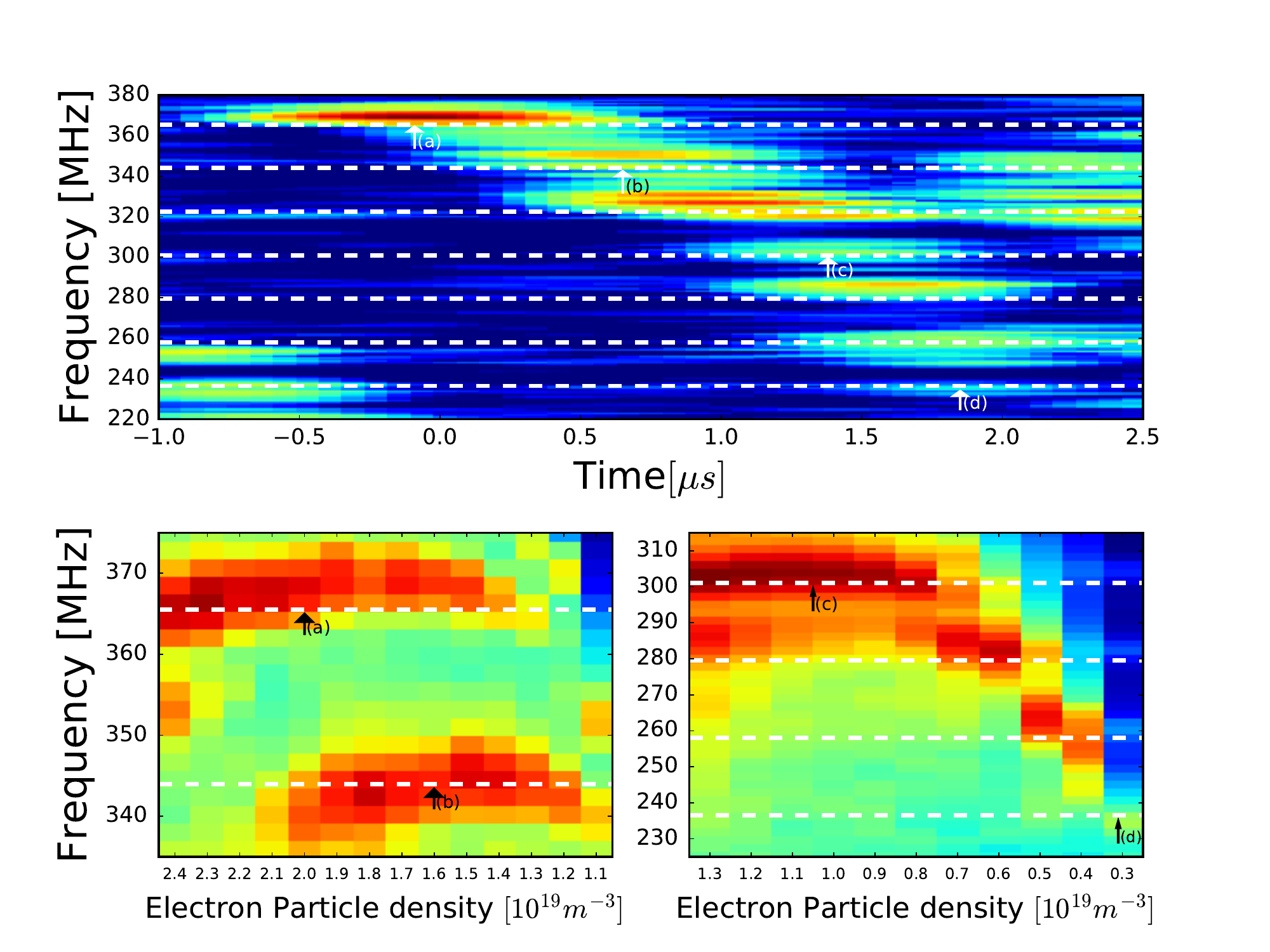}
  \vspace{-0.7cm}
\caption{\label{fig:downwardden1} Top panel: experimentally-measured fast RF burst spectrogram from KSTAR plasma 11462 with $B_{0}=1.7$T and average density before the ELM crash $\langle n_{e} \rangle = 2.5 \times 10^{19} \mathrm{m^{-3}}$. Downward step-wise frequency chirping with proton cyclotron frequency $f_{cp} \sim 21.5 \mathrm{MHz}$ is apparent. Lower panels: frequency versus density plots for the nonlinear stage of MCI simulations where $B_{z} = B_{\tiny cyc} \approx 1.41$T has been inferred from the data in the top panel. Shading indicates the $\mathrm{log_{10}}$ of the spectral power in the fluctuating part of the $B_{z}$ field component of each simulation.}
\end{figure}

Understanding the physics of edge localised modes (ELMs) \cite{elm1,elm2,elm3,elm4} in magnetically confined fusion (MCF) plasmas is crucial for the design of future fusion power plants. The same is true of the physics of the energetic ions born at MeV energies \cite{ener1, ener2} from fusion reactions between fuel ions in the multi keV thermal plasma. The crash of an ELM involves impulsive relaxation of the edge magnetic field, releasing energy and particles from the plasma at levels which may not be compatible with sustained operation of the next step fusion experiment, ITER \cite{iter1, iter2}. The confinement of fusion-born ions while they release energy, collisionally or otherwise, to the thermal plasma, was a key physics objective of the unique deuterium-tritium plasma experiments in TFTR \cite{tftr} and JET \cite{jet}, and will be central to ITER’s research programme. Here we report an unexpected conjunction of ELM physics with fusion-born ion physic. We show how this can be exploited as a diagnostic of plasma edge density with unique, very high $\left(< 1 \mu s \right)$ time resolution. This is achieved through particle orbit studies combined with first principles kinetic plasma simulations that explain high-time-resolution measurements of ion cyclotron emission (ICE) from the medium-size tokamak KSTAR \cite{kstarpre}.
We show that ICE from KSTAR deuterium plasmas is driven by a small subset of the fusion-born proton population, originating in the core of the plasma and passing through the edge region where they radiate collectively through the magnetoacoustic cyclotron instability (MCI) \cite{Dendya,Dendyb,Dendyc,McClema,Fulop,Cottrella, Schild, Cottrell,Cauff,McClem}. The MCI can occur because of the spatially localised population inversion in velocity space that is caused by the large drift excursions of 3.0$\,$MeV fusion-born protons on deep passing orbits. Our simulations of the MCI in its saturated nonlinear regime show that the frequency spectrum excited depends strongly on the plasma density. By comparing MCI spectra simulated at different densities with high-time-resolution measurements of ICE spectra during ELM crashes in KSTAR, we are able to infer the time evolution of the collapsing edge density at sub-microsecond resolution, which is unprecedented. Recently, ICE has been detected from the outer mid-plane of KSTAR \cite{Thatipamula,ayub, gunspres}, with spectral peak frequencies at local proton cyclotron harmonics; see e.g. Fig. \ref{fig:burstspec}. The only energetic protons in KSTAR plasmas are produced in deuteron-deuteron (D-D) fusion reactions, hence it is likely that ICE at spectral peaks separated by proton cyclotron harmonics is driven by fusion-born protons. If this ICE is driven by the MCI of confined fusion-born protons with spatially localised population inversion, it is necessary to identify a candidate population. The KSTAR experiment is not built on a scale sufficiently large to confine the majority of the energetic ions, 3.0$\,$MeV protons, 0.82$\,$MeV He-3 nuclei, and 1.0$\,$MeV tritons, that are born in fusion reactions within pure deuterium plasmas. To drive ICE in KSTAR via the MCI, there must nonetheless exist a collectively unstable subset of fusion protons whose orbits are confined within the plasma and traverse the excitation region in the outer midplane edge. This population must differ qualitatively from that driving ICE in JET and TFTR, because fusion-born ions on the marginally trapped orbits that give rise to ICE in JET and TFTR \cite{tftr,jet} are promptly lost from KSTAR during their first drift excursion. We identify here (see Fig. \ref{fig:orbits}) a class of drift orbits for confined D-D fusion-born ions in KSTAR that have large radial excursions and can drive the MCI. During KSTAR ELM crashes, the ICE is observed to chirp (see Fig. \ref{fig:burstspec}), often in discrete steps that equate to the local proton cyclotron frequency. Details of the emission and the detection system are in \cite{Thatipamula}. The lower panels of Figs. \ref{fig:downwardden1} and \ref{fig:downwardden2}, as well as the right panel of Fig. \ref{fig:upwardden}, provide examples of simulation results showing that the spectral structure of the fields excited by the protons depends strongly on the local plasma density. Figure \ref{fig:burstspec} shows an example of downward chirping from a pulse with toroidal magnetic field at the magnetic axis $B_0 \simeq 1.99\,$T and plasma current $I_p \simeq 600\,$kA. 

\newpage

To explore this, we have calculated orbits of 3.0$\,$MeV protons in equilibrium magnetic fields resembling those of a KSTAR plasma with major radius $R_0 = 1.8\,$m, toroidal magnetic field $B_0 = 2.27\,$T, and plasma current $I_{p}=611\,$kA. Our orbit calculations show that almost all centrally-born fusion protons are lost promptly from the plasma on their first drift orbit. However, a small fraction of these protons is born onto deeply passing confined orbits. Figure 2 shows examples of orbits of 3.0$\,$MeV protons born in the midplane at initial major radii $R(0)$ equal to (a) 1.85$\,$m and (b) 1.90$\,$m, and with initial velocity vectors slightly offset from the co-current toroidal direction. These orbits pass through the outer midplane plasma edge, and could in principle give rise to a local population inversion in velocity space, capable of driving ICE at proton cyclotron harmonics characteristic of the outer midplane. We conjecture that when the ELM crash starts, confinement of all energetic ions at the edge is lost; the edge is then rapidly re-populated on a drift orbit time-scale by the energetic protons. This leads to the sharp local population inversion that drives ICE. An instance of positive correlation between ELMs and ICE was seen on JET, see Fig. 9 of \cite{CottrellELMICE}.

To simulate the excitation of ICE by these fusion protons in KSTAR we use the EPOCH particle-in-cell code \cite{epoch}, with one spatial and three velocity dimensions. The space ($x$) direction is orthogonal to the uniform magnetic field $\bm{\mathrm{B}} = \mathrm{B_{\tiny cyc}}\, \bm{e_{z}}$, so that the propagation direction of waves excited in the system is perpendicular to $\bm{\mathrm{B}}$. $\mathrm{B_{\tiny cyc}}$ denotes the edge magnetic field inferred from the spacing between successive proton cyclotron harmonics observed in the experimental spectrograms. The bulk plasma comprises electrons and deuterons with initial temperature 1$\,$keV. Multiple simulations are carried out with initial electron densities in the range $0.2\times 10^{18}\,$m$^{-3}$ to $2.5 \times 10^{19}\,$m$^{-3}$. This range reflects Thomson scattering measurements in the edge pedestal \cite{Thatipamula}. The fusion proton population in the KSTAR plasma edge discussed above has a speed perpendicular to the magnetic field ($v_{\perp 0}$) much smaller than the birth speed. It is therefore justifiable to represent this by a delta-function ring distribution. In all our simulations, $v_{\perp 0}$ corresponds to an energy 150$\,$keV $\simeq$ 5\% of the birth energy. This is comparable to the local Alfv\'en speed $c_{A}$, hence high enough to drive the MCI \cite{dendyrev,gorerev,james,leo}, believed to generate ICE. The large parallel velocities $v_{\parallel}$ of these passing fusion protons are not represented in the simulations because for perpendicular wave propagation, the parallel dynamics of the fast ions plays no role. The ratio of proton density to deuteron density is $10^{-3}$. The total simulation duration is 10 proton cyclotron periods, which carries the MCI into its saturated nonlinear regime. Figures \ref{fig:downwardden1} and \ref{fig:downwardden2} show the spectrum of saturated ICE intensity obtained in the simulations for each initial density, along with the corresponding experimentally-measured spectrograms for downward chirping ICE during ELM crashes in plasmas with $B_{0}=1.7$T and $B_{0}=1.99$T. In the lower panels, density decreases from left to right, and each vertical strip corresponds to an independent MCI simulation at the density shown. The lower left and lower right panels correspond to the high and low frequency ranges respectively. The horizontal white dashed lines show consecutive proton cyclotron harmonics, and the arrows labelled (a)-(d) denote a mapping between experimentally observed and simulation proton cyclotron harmonics. The variation of simulated ICE intensity with density resembles the experimentally-observed variation of ICE intensity with time. There is a ``missing harmonic'' in the lower panels of Figs \ref{fig:downwardden1} and \ref{fig:downwardden2} at $f \simeq  323 \mathrm{MHz}$ and  $f \simeq  375 \mathrm{MHz}$ respectively. This could be due to periodic boundary conditions, as well as the limitations of a 1D3V model. 

\begin{figure}[!htb]
  \includegraphics[clip=true, trim = 0.5cm 0.5cm 1.0cm 1.0cm, width = \columnwidth]{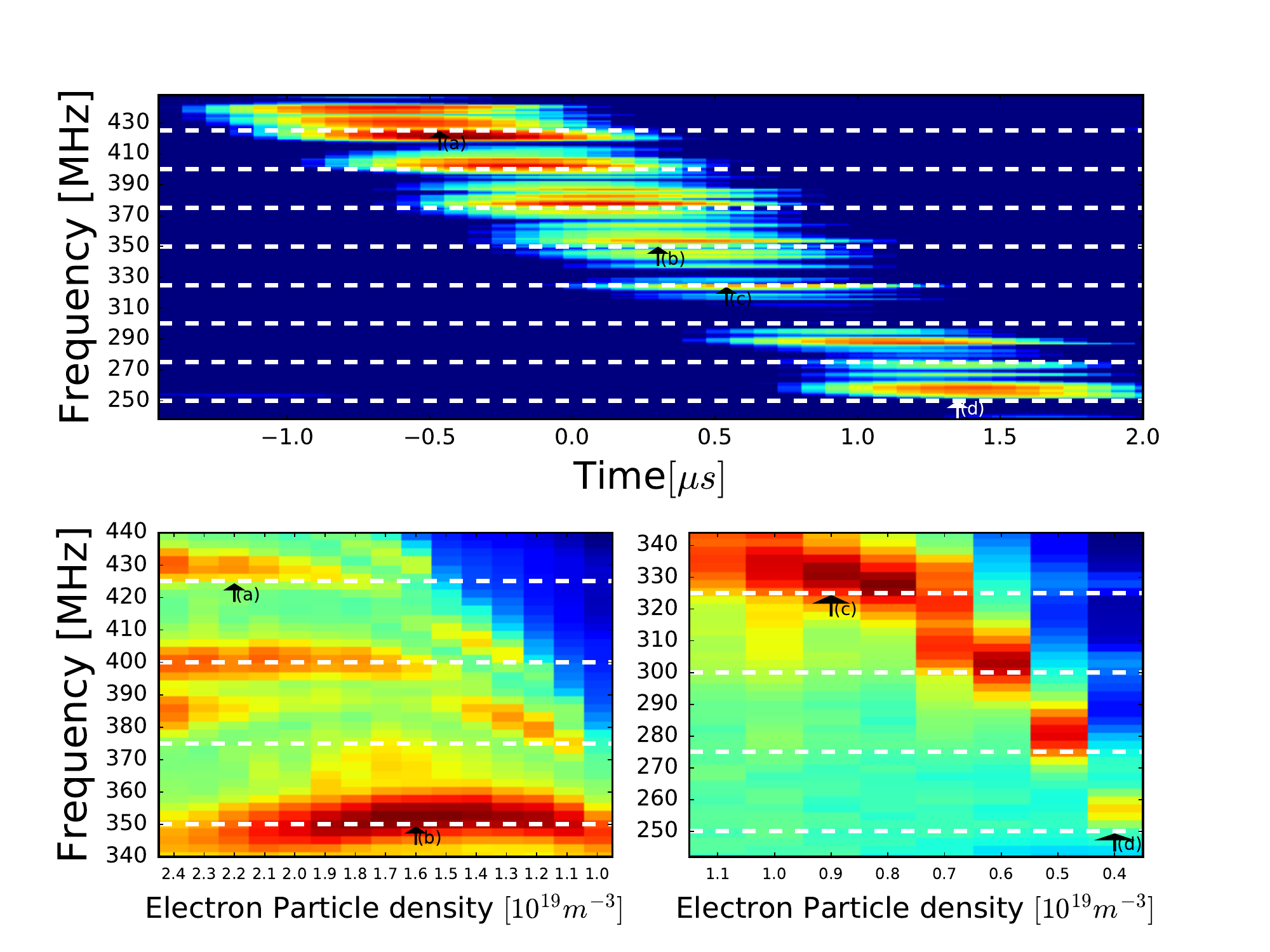}
  \label{fig:downwardden2}
    \vspace{-0.7cm}
\caption{\label{fig:downwardden2} Top panel: experimentally-measured fast RF burst spectrogram from KSTAR plasma 11513 with $B_{0}=1.99$T and average density before the ELM crash $\langle n_{e} \rangle = 2.6 \times 10^{19} \mathrm{m^{-3}}$. Downward step-wise frequency chirping with $f_{cp} \sim 25 \mathrm{MHz}$ is apparent. Lower panels: as the lower panel of Fig. \ref{fig:downwardden1} but with $B_{z} = B_{\tiny cyc} \approx 1.64$T.}
\end{figure}

\begin{figure}[!htb]
  \includegraphics[clip=true, trim = 1.0cm 7.2cm 1.0cm 1.0cm, width = \columnwidth]{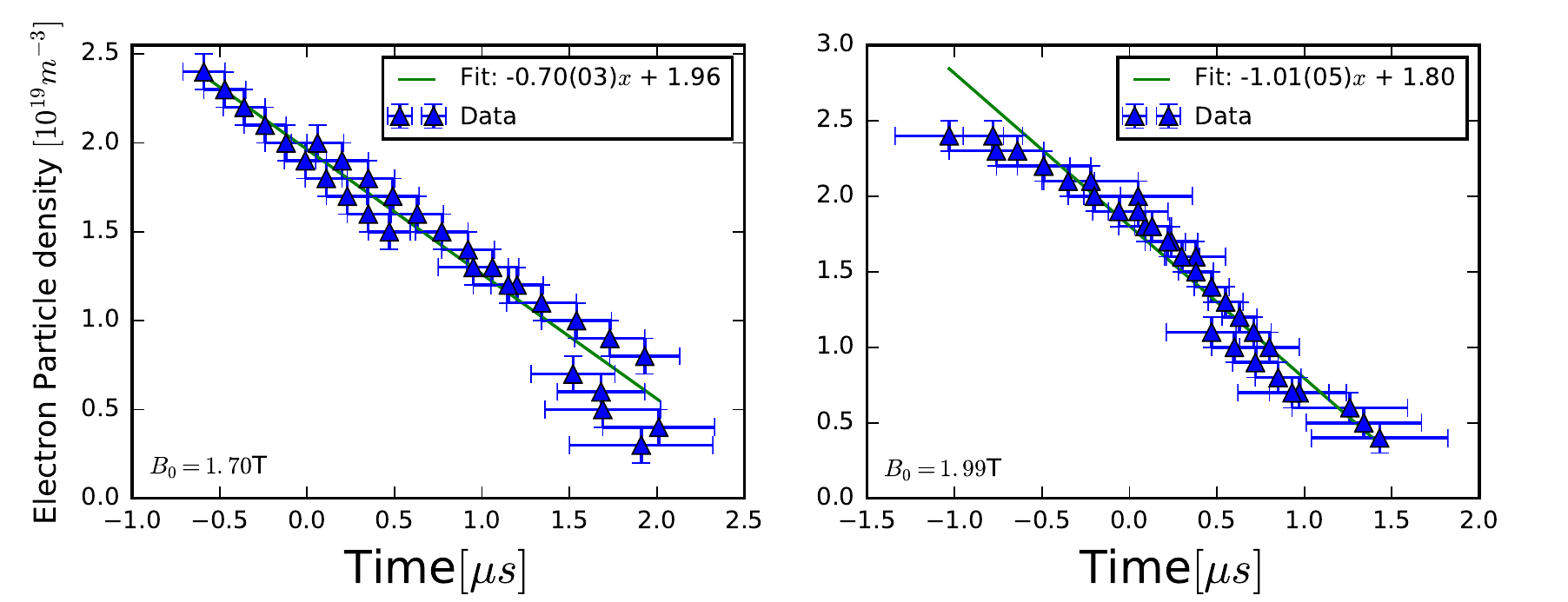}
    \vspace{-0.8cm}
\caption{\label{fig:deneqdown} Pedestal density during an ELM crash in KSTAR inferred from downward chirping ICE measurements compared with saturated MCI simulations. Left (right): fitted and estimated density and time values corresponding to KSTAR plasma 11462 (plasma 11513). The green fit suggests that local density declines approximately linearly with time during the early stages of the ELM crash.}
\end{figure}

It is known that the density at the top of the edge pedestal in an H-mode plasma collapses during an ELM crash. Our simulations (see also Fig. \ref{fig:downwardden2}) show that the downward chirping of ICE observed during such a crash is likely to be a direct consequence of the density collapse. An important corollary of this result is that measurements of the ICE spectra can be used to infer the time-evolving density in the pedestal. This is shown in Fig. \ref{fig:deneqdown}, the left panel corresponding to Fig. \ref{fig:downwardden1}, and the right panel corresponding to Fig. \ref{fig:downwardden2}. In this plot, error bars in time reflect the uncertainty in determining the start and end times of each cyclotron harmonic feature in the experimental plots. Density error bars are due to the finite steps in density between different simulations. For Fig. \ref{fig:deneqdown}, the time resolution of the density collapse is of order $\sim 0.1 \mu s$. Overall, Fig. \ref{fig:deneqdown} shows that the density collapses approximately linearly on a timescale of about 2.5$\,\mu$s.   

\begin{figure}[!htb]
  \includegraphics[clip=true, trim = 0.5cm 7.2cm 1.0cm 1.0cm, width = \columnwidth]{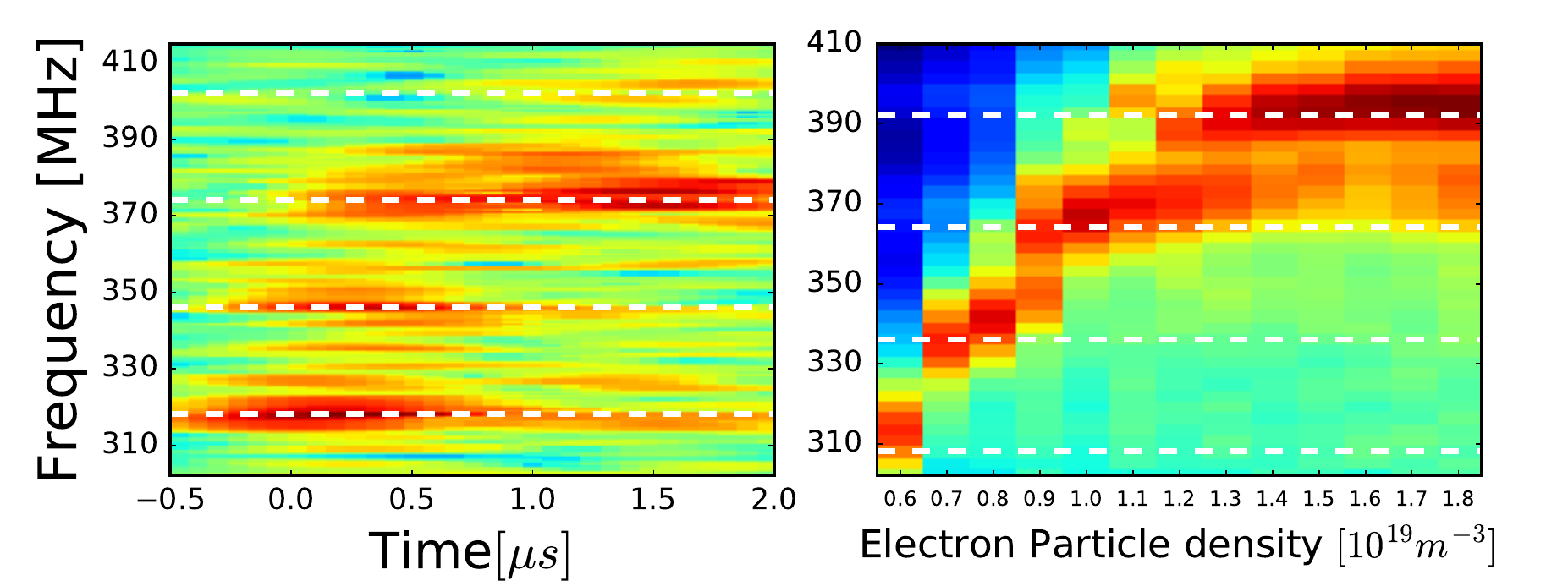}
    \vspace{-0.7cm}
\caption{\label{fig:upwardden} Left panel: experimentally-measured fast RF burst spectrogram from KSTAR plasma 11474 with $B_{0}=2.27T$ and $\langle n_{e} \rangle = 2.5 \times 10^{19} \mathrm{m^{-3}}$. Upward step-wise frequency chirping with $f_{cp} \sim 28 \mathrm{MHz}$ is apparent. Right panel: as the lower panel of Fig. \ref{fig:downwardden1} but with $B_{z} = B_{\tiny cyc} \approx 1.84$T.}
\end{figure}

\begin{figure}[!htb]
  \includegraphics[clip=true, trim = 1.0cm 7.2cm 1.0cm 1.0cm, width = \columnwidth]{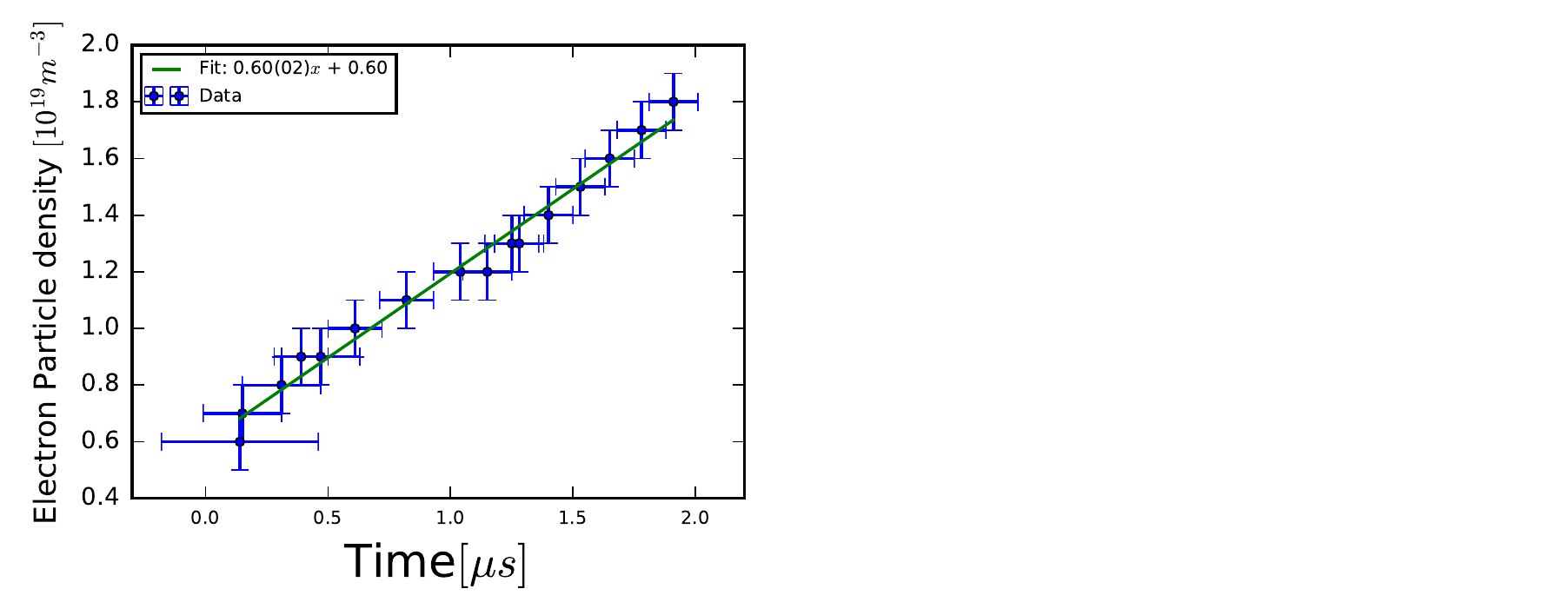}
      \vspace{-0.8cm}
\caption{\label{fig:denequp} Rising density during an ELM crash inferred from upward chirping ICE measurements in KSTAR plasma 11474 combined with saturated MCI simulations at different densities, see Fig. \ref{fig:upwardden}. The green fit suggests suggests density rises approximately linearly with time during the early stages of the ELM crash.}
\end{figure}

Upward chirping is also sometimes observed during ELM crashes in KSTAR. This can also be explained in terms of our model if, in these cases, ICE originates from regions of locally rising edge density associated with ELM filaments. The time evolution of the corresponding density for the case of upward chirping shown in Fig. \ref{fig:upwardden} (left panel) has been inferred from the saturated MCI simulations at different densities (Fig. \ref{fig:upwardden} right panel), and is shown in Fig. \ref{fig:denequp}. 

In Fig. \ref{fig:upwardden}, although the separation of ICE spectral peaks is $f_{cp}$, the peaks differ systematically from integer multiples of $f_{cp}$ by $\sim 10$MHz. This discrepancy might be due to a Doppler shift arising from the rapid motion of an ELM filament. 

\newpage

We have shown that harmonic ICE with spacing equal to $f_{cp}$ in KSTAR deuterium plasmas is driven by a small subset of the fusion-born proton population originating in the core of the plasma. The drift orbits of these protons have large radial excursions to the outer midplane edge. We have compared the nonlinearly saturated field spectra obtained from multiple MCI simulations at different plasma densities with experimentally observed time evolving ICE spectra. By combining different simulation spectra with the chirping ICE observed during KSTAR ELM crashes, we obtain a sub-microsecond time resolution diagnostic of the plasma density in the emitting region. Downward chirping ICE predominates observationally, and can be used to quantify the collapse of edge density during the ELM crash. Upward chirping ICE, which is occasionally seen in KSTAR, may originate from ELM filaments. Combined with the passive, non invasive character of ICE measurements, the results of this paper suggest an attractive way forward for future energetic ion measurements in ITER \cite{iter1, iter2}.
%regions of locally rising edge density associated with

% acknowledgement_paragraph_r2.tex                25 June 2014
%
The authors thank J W S Cook for discussions. This project used the EPOCH code, part funded by UK EPSRC grants EP/G054950/1, EP/G056803/1, EP/G055165/1 and EP/ M022463/1. This work received funding from the RCUK Energy Programme [grant number EP/I501045], Euratom, and the National Research Foundation of Korea [grant number 2014M1A7A1A03029881]. The views and opinions expressed herein do not necessarily reflect those of the European Commission.
%
   % input acknowledgement

\end{document}